\documentclass[prb,twocolumn,showpacs,preprintnumbers,superscriptaddress,aps,longbibliography]{revtex4-1}

\usepackage{amssymb}
\usepackage{color}

\def \Prague{Institute of Physics, Charles University, Faculty of Mathematics and Physics, Ke Karlovu 5, Prague 2, CZ-121~16, Czech Republic}

\usepackage{epsfig}
\usepackage{amsmath}

\begin{document}
\title{Intercalation of hydrogen in the SiC/epitaxial graphene interface}

\author{J. \surname{Kunc}}
\email{kunc@karlov.mff.cuni.cz}
\author{M. \surname{Rejhon}}
\author{P. \surname{Hl\'{i}dek}} \affiliation\Prague

\date{\today}

\begin{abstract}
We have measured optical absorption in mid-infrared spectral range on hydrogen intercalated epitaxial graphene grown on silicon face of SiC. We have used attenuated total reflection geometry to enhance absorption related to the surface and SiC/graphene interface. The samples of epitaxial graphene have been intercalated in the temperature range of 790 to 1250$^\circ$C and compared to the reference samples of hydrogen etched SiC.
We have found that although the Si-H bonds form at as low temperatures as 790$^\circ$C, the well developed bond order has been reached only for epitaxial graphene intercalated at temperatures exceeding 1000$^\circ$C. We also show that the hydrogen intercalation degradates on a time scale of few days when samples are stored in ambient air.
The optical spectroscopy shows on a formation of vinyl and silyl functional groups on the SiC/graphene interface due to the residual atomic hydrogen left from the intercalation process.
\end{abstract}


\maketitle

\section{Introduction}
Graphene as a semi-metal with tunable Fermi level is an alternative concept to modify Schottky or tunneling barriers formed at the graphene/semiconductor or graphene/oxide/metal interface. The top gated epitaxial graphene grown on SiC(000$\bar{1}$)~\cite{JouaultAPL2012} and SiC(0001)~\cite{ShenAPL2009,TanabeAPL2010,MoonGaskillIEEE2010} has been demonstrated. A back-gated epitaxial graphene provides direct access to the graphene/semiconductor interface~\cite{WaldmannNatureMat2011,WaldmannJPDApplPhys2012} and it would facilitate optical studies without complications caused by underlying substrate~\cite{FrommNJP2013}. Bottom-gated epitaxial graphene can be also used to reduce carrier scattering caused by top-gate~\cite{PulsAPL2011}, or it can be used in tunable single molecule transistors~\cite{UllmannWeber2015}. However, backgating of epitaxial graphene grown on silicon face of SiC(0001) has been demonstrated only on hydrogen intercalated graphene~\cite{WaldmannNatureMat2011,WaldmannJPDApplPhys2012}.

The as---grown epitaxial graphene on SiC(0001) consist of so called zero graphene layer, also called buffer layer, and the single layer graphene. The buffer layer is known to contain roughly 30\%~\cite{EmtesvPRB2008} of sp$^3$ bonded carbon. These carbon atoms are bonded to Si in SiC beneath the buffer layer. Due to the low degree of order of these sp$^3$ bonds, originating from $6\sqrt{3}\times6\sqrt{3}R30^\circ$ SiC surface reconstruction, the band structure of buffer contains large amount of localized states~\cite{MattauschPRL2007}. These interface localized states pin the Fermi level when as---grown epitaxial graphene is gated. Another issue is that the buffer layer mediates interaction between carriers in the graphene layer and phonons in SiC, thus significantly reducing carrier mobility in graphene~\cite{GiesbersPRB2013,YuAPL2013,LinAPL2014}. Therefore, the high mobility graphene with tunable Fermi level requires eliminating buffer~\cite{RobinsonNanoLetters} and thus reducing interaction between graphene and SiC~\cite{RiedlPRL2009,FortiPRB2011}. This can be readily done by hydrogen intercalation~\cite{RiedlPRL2009} in the SiC/buffer interface, where hydrogen saturates Si-C bonds, turns sp$^3$ carbon back into sp$^2$ bonded carbon and, as a result, so called buffer-free quasi free standing bilayer graphene is formed~\cite{RiedlPRL2009,CuiAPL2014}.

Beside hydrogen, also annealing in oxygen~\cite{KimArxivOxygen2017}, rapid cooling~\cite{BaoNorimatsuPRL2016} or ion implantation~\cite{StoehrPRB2016} have been shown to turn buffer layer into quasi free standing single layer graphene. However, oxygen annealing and ion implantation can lead to defect formation in graphene and rapid cooling is not suitable for fabrication of electronic devices on large scales. The best solution still seems to be hydrogen intercalation since it is process readily  available in semiconductor industry. 

The main objective of this work is to study Si-H bond formation and its temporal stability. This will provide a route towards reliable, back-gated epitaxial graphene of high carrier mobility. Although temperature of Si-H bond formation is well known (710$^\circ$C), it is not clear how this temperature changes in the case of SiC/buffer interface and how well-ordered the Si-H bonds are. In the case of SiC/single layer graphene (SLG) interface, hydrogen has to either tunnel through the single layer graphene and buffer layer, or, it gets in the SiC/buffer interface via defects in graphene and/or buffer. Therefore we assume that the temperature of fully decoupled buffer from SiC is rised at fixed hydrogen intercalation times. The temporal stability of intercalated hydrogen is also investigated here and we discuss potential issues caused by residual atomic hydrogen in the SiC/graphene interface. 

\begin{figure}[t!]
\centering
\includegraphics[width=9cm]{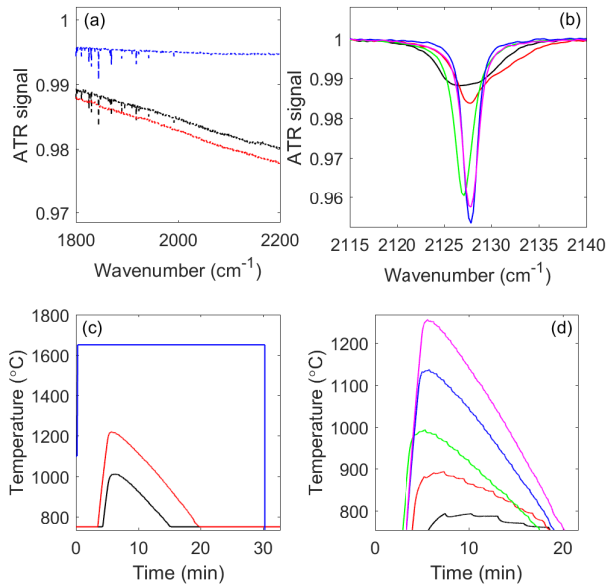}
\caption{The ATR spectra of H$_2$ treated (a) bare SiC and (b) single layer graphene on SiC. The growth regimes (c) for H$_2$ treated bare SiC and (d) single layer graphene on SiC. The ATR spectra in (a) and growth recipes in (c) of bare SiC samples annealed at 1650, 1210, 1010$^\circ$C are depicted by blue, red and black curves, respectively. The ATR spectra in (b) and growth recipes in (d) of single layer graphene samples annealed at 1250, 1140, 990, 890 and 790$^\circ$C are depicted by magenta, blue, green, red and black curves, respectively.}
\label{Fig1}
\end{figure}
\begin{figure}[t!]
\centering
\includegraphics[width=7cm]{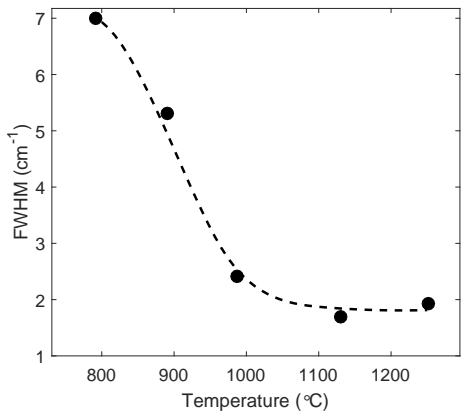}
\caption{The FWHM of the Si-H absorption band in H$_2$ intercalated single layer graphene versus maximal temperature of H$_2$ intercalation. The full annealing recipes are shown in Fig.~\ref{Fig1}~(d).}
\label{FigFWHM}
\end{figure}

\section{Experimental results}
The 4$H$-SiC wafers were bought from II-VI Inc. We use on-axis ($\pm0.6^\circ$) semi-insulating SiC, 500~$\mu$m thick wafers with resistivity $\rho>10^{9}~\Omega$cm. The conducting nitrogen doped SiC wafers have resistivity $\rho=15-28\times10^{-3}~\Omega$cm, thickness 350~$\mu$m and they are cut 4$^\circ$ off the c-axis (0001). We grow epitaxial graphene on epi-ready (chemically mechanically polished) Si-face. The wafers are diced on 3.5$\times$3.5~mm$^2$ samples. The epitaxial graphene is grown in inductively heated furnace~\cite{deHeerPNAS2011} at 1600$^\circ$C for 5~minutes in argon atmosphere and argon flow 30~standard liters per hour (SLPH). More details about the growth conditions can be found in our previous work~\cite{KuncPRAppl}. Annealing in hydrogen is performed in the range of temperatures from 790$^\circ$ to 1650$^\circ$C. The hydrogen pressure and flow are kept at 1000~mbar and 30~SLPH, respectively. 
The epitaxial graphene samples are characterized by micro-Raman confocal spectroscopy. As Si-H bond is formed only at the SiC/graphene interface, the measurements in transmission geometry are precluded by high background signal from bulk SiC. Therefore it is essential to measure by means of Attenuated Total Reflection (ATR)~\cite{SpeckAPL2011}, employing evanescent wave of totally reflected light which probes only the SiC surface layer. The probed surface layer thickness is few micrometers, still four orders of magnitude more than the thickness of Si-H layer. In order to surpass the obstacle of high refraction index of SiC in mid-infrared spectral range we use germanium crystal which has comparably high index of refraction as SiC. The ATR spectra are measured by evacuated Fourier Transform Infrared (FTIR) spectrometer Bruker Vertex80v. The ATR module from Pike Technologies is equipped by single-reflection germanium crystal and the sample is placed on top of the crystal and pressed by calibrated pressure clamp.  

We compare H$_2$ annealing of bare SiC, Fig.~\ref{Fig1}~(a) and single layer epitaxial graphene  (SLG), Fig.~\ref{Fig1}~(b). We have annealed bare SiC at temperatures up to 1210$^\circ$C and the annealing procedures are the same as for H$_2$ intercalation of SLG, see Fig.~\ref{Fig1}~(c), red and black curves for growth recipes. There is no signature of Si-H absorption band in H$_2$ annealed SiC. No signature of Si-H bond is observed even after high temperature annealing at 1650$^\circ$C for 30~minutes, as shown in Fig.~\ref{Fig1}~(a) by blue ATR spectrum. However, well pronounced Si-H absorption band is developed at 2128~cm$^{-1}$ in SLG samples, as shown in Fig.~\ref{Fig1}~(b). The annealing temperature during H$_2$ intercalation is depicted in Fig.~\ref{Fig1}~(d). The presence of Si-H absorption band on SLG samples and the absence of Si-H absorption band in bare SiC samples proves that the Si-H bond is formed at the SiC(0001)/SLG interface and it is not present in the bulk of SiC. We observe also that the Full Width at Half Maximum (FWHM) of Si-H absorption band narrows as the temperature of H$_2$ intercalation increases. We depict the FHMM versus maximal temperature of H$_2$ intercalation in Fig.~\ref{FigFWHM}. The H$_2$ intercalation at temperatures as low as 790$^\circ$C leads to relatively weak and broad absorption band. As the temperature is rised the FWHM of the Si-H band narrows and the FWHM saturates at $\delta\omega=(1.8\pm0.2)$~cm$^{-1}$. We note that the ATR spectra are measured with FTIR spectrometer at the spectral resolution 0.5~cm$^{-1}$, hence the saturation value $\delta\omega$ is not caused by the minimal spectral resolution of the FTIR spectrometer.

\begin{figure}[t!]
\centering
\includegraphics[width=8cm]{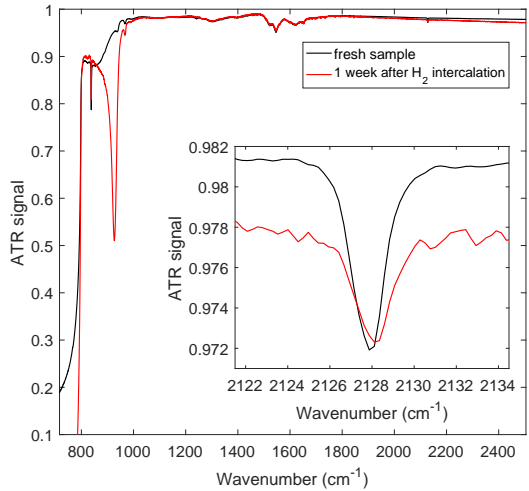}
\caption{ATR spectra of single layer graphene sample measured (black curve) immediately after H$_2$ intercalation and (red curve) one week after the annealing. The sample was stored in air. (inset) Detail of Si-H absorption band as measured (black curve) immediately after H$_2$ intercalation and (red curve) one week after the annealing. }
\label{FigTimeEvolution}
\end{figure}

\begin{figure}[t!]
\centering
\includegraphics[width=8cm]{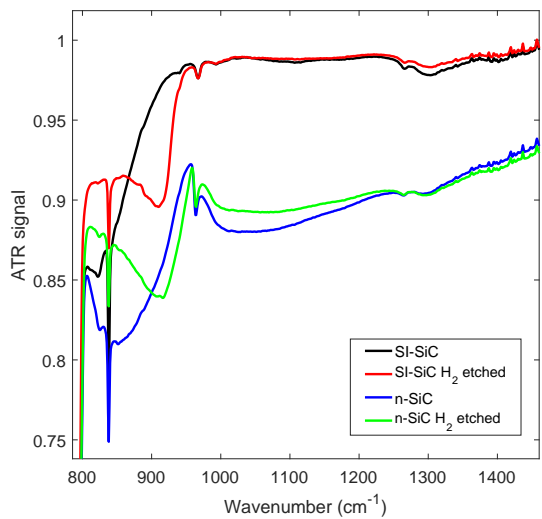}
\caption{ATR spectra of SiC (black and blue curves) before H$_2$ annealing and (red and green curves) after H$_2$ annealing. The black and red (blue and green) curves correspond to semi-insulating (conducting) SiC.}
\label{FigH2etchingofSiC}
\end{figure}

We study also temporal stability of H$_2$ intercalation of SLG. The Si-H bond is well pronounced in the SLG sample measured immediately after H$_2$ intercalation, as shown by black curve in the inset of Fig.~\ref{FigTimeEvolution}. When sample is left in ambient air for one week, the intensity of Si-H absorption band decreases by 50\% and its FWHM increases by $4$\%, as shown by red curve in the inset of Fig.~\ref{FigTimeEvolution}. This is evidence of gradually reduced amount of Si-H bonds. The second effect of aging is a newly developed band at 900-940~cm$^{-1}$. This band can be also formed without aging, as we show in Fig.~\ref{FigH2etchingofSiC}. The band at 900-940~cm$^{-1}$ is formed in bare SiC when we etch the sample at 1650$^\circ$C for 30~minutes. We observe similar behavior for both semi-insulating and conducting bare SiC samples, as we show in Fig.~\ref{FigH2etchingofSiC}. We note that this band is not observed when annealing temperature of SiC is below 1200~$^\circ$C.

\section{Discussion}
We have shown presence of Si-H chemical bonds in hydrogen intercalated single layer epitaxial graphene. The H$_2$ intercalation leads to quasi-free standing bilayer graphene as has been discussed in literature and confirmed by our Raman spectroscopy (data not shown here). The bond order can be deduced from the FWHM of 2128~cm$^{-1}$ Si-H absorption band. The observed broadening of this band can be interpreted as lower degree of bond order with respect to the direction perpendicular to the sample surface. The lowest temperature of Si-H bond formation at T$_{min}=700^\circ$C corresponds to the onset of Si-H absorption band is SLG samples at T$_{min,SLG}=790^\circ$C. This low temperature Si-H bond formation is however not complete as shown by FWHM as large as 7~cm$^{-1}$. The well ordered Si-H bonds develop at temperatures 300$^\circ$C higher (above 1000$^\circ$C). We interpret this temperature rise as a consequence of additional barrier which hydrogen has to overcome to get in SiC/SLG interface. The two possible routes are defects in graphene and buffer, or, since hydrogen is small molecule, it can penetrate through the graphene layer. \\
Beside the successful H$_2$ incorporation in the SiC/SLG interface, it is also important that the formed intercalated SLG is stable for reliable device operation. Our observation of degradation of H$_2$ intercalation shows that the reliability needs to be addressed in the following research. Beside our observation that amount of Si-H bonds is reduced by 50\% in one week, we have also observed formation of new absorption band at 890-940~cm$^{-1}$. As we are able also to form this band by annealing SiC at high temperature (1650$^\circ$C) for 30~minutes, we argue that it should be related to either CH or SiH groups formed on the SiC surface. As the FWHM of this absorption band reaches 50~cm$^{-1}$, it is probably composed of many different types of bonds and configurations. We suggest that energetically these chemical groups could be SiH$_2$ (900-930~cm$^{-1}$)~\cite{MasudaJMCC2015}, deformation mode of SiH$_3$ (920, 843.5, 939.6, 941~cm$^{-1}$)~\cite{SavocaPCCP2013}, Si-H bending (937~cm$^{-1}$)~\cite{SchmidtChemMat1991}, Si-H$_2$ scissor mode (910~cm$^{-1}$)~\cite{XuNJC2003} or vinyl groups (880-950~cm$^{-1}$)~\cite{SkibbeJCP2008}. Monosubstituted alkenes have a vinyl group with fundamental modes at 911~cm$^{-1}$ for CH wagging and 925~cm$^{-1}$, CH$_2$ wagging mode~\cite{NaritaJCP1959}, other possibilities are geminal disubstituted alkene $\mathrm{R_2C=CH_2}$ (880-900~cm$^{-1}$) or $\mathrm{-SiCH_3}$ (750-900~cm$^{-1}$ and 1240-1275~cm$^{-1}$)~\cite{MajumdarJES2008}. The time evolution of infrared absorption suggests that hydrogen tends to further react with silicon or carbon in the buffer/SLG layer. This is probably atomic hydrogen left below the buffer layer from the hydrogen intercalation. Mono- and disubstituted alkenes can be understood as a step towards gaseous silane, methane and acetylene formation, as has been shown experimentally by mass spectrometry where atomic hydrogen reacts with SiC even at room temperature~\cite{KimOlanderSurfSci1994}. We note that although we intercalate SLG by molecular hydrogen, the hydrogen molecule is dissociated prior to bonding to silicon in the process of buffer decoupling. Inevitably, the second hydrogen atom is left in the vicinity of the Si-H bond formed by the first hydrogen atom. The high reactivity of atomic hydrogen causes formation of methane, acetylene and silane, creating defects in the quasi-free standing bilayer graphene.

\section{Conclusions}
Hydrogen intercalation is an essential step towards back-gated epitaxial graphene, the key prerequisite to employ graphene unique property as a metal with tunable Fermi energy. We have shown optimal conditions for hydrogen intercalation of the single layer graphene. We have also shown that the stability of hydrogen intercalated SLG is largely reduced due to the presence of highly reactive atomic hydrogen, which is left behind Si-H bond formation during decoupling of the buffer layer from SiC substrate. This reaction of atomic hydrogen tends to form vinyl and silyl functional groups on the SiC/SLG surface, as shown by infrared absorption spectroscopy. 

\section{Acknowledgement}
Financial support from the Grant Agency of the Czech Republic under project 16-15763Y, and the project VaVpI CZ.1.05/4.1.00/16.0340 are gratefully acknowledged.

\end{document}